# SVM Classifier on Chip for Melanoma Detection

Shereen Afifi[1], Hamid GholamHosseini[2], and Roopak Sinha[3]
[1&2]Electrical and Electronic Engineering Department, Auckland University of Technology
Auckland 1010, New Zealand
[3]Department of IT and Software Engineering, Auckland University of Technology
Auckland 1010, New Zealand
e-mails: [1]safifi@aut.ac.nz, [2]hgholamh@aut.ac.nz, [3]rsinha@aut.ac.nz

**Abstract**

*Support Vector Machine (SVM) is a common classifier used for efficient classification with high accuracy. SVM shows high accuracy for classifying melanoma (skin cancer) clinical images within computer-aided diagnosis systems used by skin cancer specialists to detect melanoma early and save lives. We aim to develop a medical low-cost handheld device that runs a real-time embedded SVM-based diagnosis system for use in primary care for early detection of melanoma. In this paper, an optimized SVM classifier is implemented onto a recent FPGA platform using the latest design methodology to be embedded into the proposed device for realizing online efficient melanoma detection on a single system on chip/device. The hardware implementation results demonstrate a high classification accuracy of 97.9% and a significant acceleration factor of 26 from equivalent software implementation on an embedded processor, with 34% of resources utilization and 2 watts for power consumption. Consequently, the implemented system meets crucial embedded systems constraints of high performance and low cost, resources utilization and power consumption, while achieving high classification accuracy.*

## 1. INTRODUCTION

Melanoma is the most dangerous form of skin cancer worldwide with highest rates in New Zealand and Australia. Early diagnosis of melanoma would aid in reducing mortality rates as well as the treatments costs. Recently, skin cancer specialists are using Computer Aided Diagnosis (CAD) systems as a decision support tool for early detection of melanoma. Thus, a real-time embedded CAD system in the form of a low-cost handheld device dedicated for melanoma detection is needed in the primary care. However, developing such embedded system is so challenging because of its complicated computations. Field-Programmable Gate Array (FPGA) is a powerful parallel processing reconfigurable device that is widely used for achieving essential performance of embedded systems, while effectively utilizing hardware resources, offering low cost and low power consumption.

Interestingly, FPGAs have recently demonstrated significant performance in various applications and outperform other comparable platforms [1]. Hence, FPGA is an ideal platform for developing a real-time embedded CAD system for melanoma detection with high performance and low cost.

Our research group has developed software algorithms to implement the four stages of the CAD structure (Fig. 1), targeting efficient melanoma detection [2]. The final classification stage in the CAD structure was concluded to be the most compute-intensive stage, which needs hardware/FPGA acceleration. The Support Vector Machine (SVM) classifier was tested for melanoma detection and verified based on experimental results that shows higher classification accuracy compared to other tested classifiers. Also, SVM is a powerful supervised machine learning tool that demonstrates high classification accuracy for various applications. Therefore, this research study focuses on implementing and accelerating SVM onto FPGA to be embedded within a cost-effective handheld device running a CAD system for early melanoma detection.

Different hardware implementations of SVM classifier exploit the parallel processing power of FPGAs for achieving high performance computing [3]. Some hardware implementations recorded relatively loss in the classification accuracy rate compared to software implementations. Most implementations are realized on old FPGA platforms with traditional methods. Furthermore, many architectures are developed without taking into consideration important embedded systems constraints like low power consumption that was measured for only a few numbers of previous implementations. Finally, it was concluded that the main challenges are the difficulty of meeting important embedded systems constraints of high performance, flexibility, scalability, and low area, cost, and power consumption, while reaching reliable classification with high accuracy. In this paper, we implement optimized FPGA-based SVM classifier and cascaded classifier using the latest platform and design methodology targeting melanoma detection. Experimental results demonstrate meeting embedded systems constraints, while achieving accurate classification on a single device/chip.

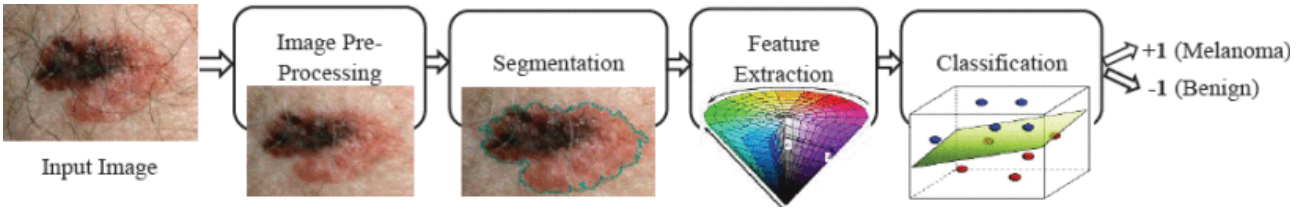

Figure 1. The CAD system structure.

## 2. PROPOSED SVM DESIGN

**A. FPGA Platform and HLS Design Methodology**

The latest FPGA technology and system-design tool are selected for our implementation. The Xilinx "Zynq-7000 All Programmable System on Chip (SoC)" is a recent FPGA platform, which simplifies the embedded system design and realization due to its hybrid architecture [4]. The hybrid structure of the Zynq SoC combines the hardware programmability of an FPGA as a Programmable Logic (PL) with an ARM Cortex-A9 dual-core as a Processing System (PS) in a single SoC.

The recent version of the Xilinx "Vivado Design suite" is used for the design and development process. The Vivado suite includes the HLS tool that employs the latest UltraFast HLS design methodology [5]. The HLS methodology is discriminated by using the high-level language instead of the traditional hardware description language for programming the FPGA. By using the HLS, the hardware development effort and time-to-market are significantly reduced.

**B. Proposed SVM HLS IP**

This proposed design of an SVM IP is based on our previous implementation [6] to be extended and improved. The HLS design methodology is used to implement a binary SVM classifier as an HLS IP. The SVM IP basically implements the main classification function (1) that is used in the classification phase for classifying any new test instance, $X$, based on the sign function. The linear kernel is applied, which performs the complicated dot-product calculation between the test sample $X$ and each Support Vector (SV) as $X_i$. The function is based on the number of $SV$ (as $N$) and it uses other parameters for the required calculation, $\alpha$, $y$ and $b$ that are identified from the training phase.

$$F(X) = sign\left(\sum_{i=1}^{N} \alpha_i y_i (\vec{X} \cdot \vec{X_i}) - b\right) \quad (1)$$

The initial hardware/software co-design for an accelerator IP in [6] is to be extended in this paper to implement the whole function (1) to realize an optimized SVM IP running on the recent Zynq SoC for fast, low-cost, and reliable melanoma detection. Using the HLS tool, a module is designed in C/C++ implementing the classification function (1) to develop an SVM HLS IP. The proposed module has three main inputs as arrays and is divided into three functional blocks to compute the required calculations.

The block diagram of the proposed IP is shown in Fig. 2. Three input ports are designed for the arrays, which are called according to its contents as $SVs$, $Parameters$, and $X$ arrays. The $SVs$ is a 2D array that contains the features data of each SV. The $parameters$ and $X$ are 1D arrays that hold $\alpha y$ values of each SV (and the $b$ value), and features of the test instance $X$ respectively. Apparently, the proposed design depends on both the number of SVs and features size to implement any SVM model. To simplify the design, the function calculation is divided into three main successive blocks. The first block "SVs Summation" is designed in a nested for loop (with the size of SVs number times features number) to compute the first part of the summation (accumulated array/vector Z) in the main function as in (2).

$$\vec{Z_i} = \sum_{i=1}^{N} \alpha_i y_i \vec{X_i} \quad (2)$$

The second block "Distance Calculation" performs the dot-product calculation between the accumulated array $Z$ that outputted from the first block and the test array $X$ as in (3) to produce the classification distance value $D$.

$$D = \sum_{i=1}^{N} \vec{X} \cdot \vec{Z_i} \quad (3)$$

Finally, the parameter $b$ (stored in the first place of the $parameters$ array) is subtracted from the distance $D$ to be then classified according to the proposed sign function in (4) at the final block "Classification Decision" ($th$ is the threshold value determined through the validation phase). The final classification output $F(X)$ is returned through the control bus/interface $Ctrl$, which corresponds to a melanoma class (+1), benign/non-melanoma class (−1).

$$F(X) = sign(D - b) = \begin{cases} -1, & (D - b) < th \\ 1, & (D - b) \geq th \end{cases} \quad (4)$$

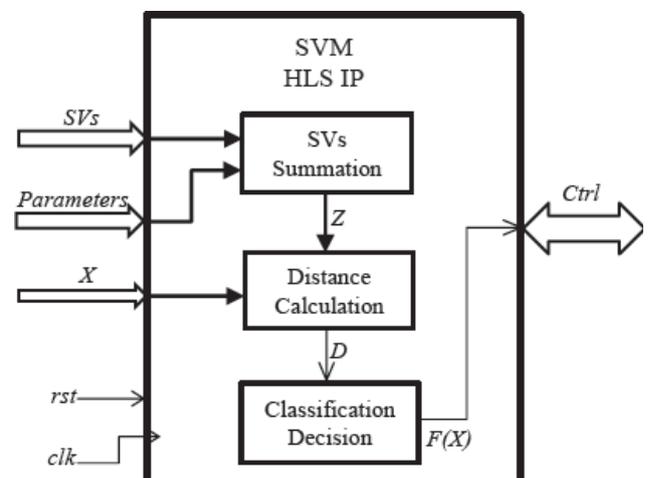

Figure 2. Proposed design of the SVM HLS IP.

The HLS tool simplifies hardware design by assigning interfaces, resources, and other hardware techniques to the module through available various directives. The three input ports are each assigned as a BRAM interface (directive), as well, each input array is mapped to a single-port RAM. The control bus is allocated to an AXI-lite bus, which controls the designed IP and data flow of the system through communicating with the ARM processor in the Zynq PS part.

In addition, the pipelining technique is applied for each loop in order to improve the data throughput and latency. Accordingly based on the HLS synthesis results for an SVM model with 248 SVs and 27 features, the latency is reduced from 67,294 clock cycles to **8091** clock cycles, gaining a speedup of 8x from applying loop pipelining, while same number of DSPs (5) is utilized. As a disadvantage of gaining significant acceleration, extra FF and LUT resources are utilized, which reflects the well-known trade-off between speed and area. However by using the modern Zynq SoC of extensive resources, low figures of only 17% (18941) and 36% (19308) are utilized for FFs and LUTs respectively, while gaining significant speed improvement. Consequently, the designed pipelined HLS IP is promising for achieving HPC, and low area, cost and power consumption requirements of embedded systems realization for enhancing melanoma detection.

The designed HLS IP is successfully implemented and packed as an IP/RTL implementation by passing through the design flow of the HLS tool (C simulation, C synthesis, RTL co-simulation, RTL implementation). Then, the implemented HLS IP is integrated in the proposed system as depicted in Fig. 3 that is designed by exploiting the Vivado design tool to be realized on Zynq SoC.

The ARM processor in the Zynq PS part is connected to the HLS IP and other used IPs/cores in the Zynq PL via the control bus (AXI-lite) using an AXI-interconnect IP, in order to control connected IPs and data flow in the system. Three dual-port BRAMs are instantiated to be connected to the HLS IP input ports on one port each and to the ARM on the second port using AXI-BRAM-Controller IPs. In addition, an AXI-Timer IP is connected to the system to measure the clock cycles needed by running the connected IPs, which is exploited for performance comparisons.

The designed SoC in Fig. 3 is successfully implemented via the Vivado design tool stages (synthesis, implementation and bitstream generation) to be finally exported for running on Zynq by the help of the Software Development Kit (SDK) tool. A test bench (C program) is developed for testing and verifying the implemented classifier and run it on Zynq SoC. The Zynq ARM processor executes the developed program/application besides controlling the system. The data required for the three arrays inputs of the HLS IP is read Byte-by-Byte from three main files to be parsed and then written into the corresponding BRAMs. The files are stored in the SD card of the Zynq evaluation board, which hold required data for running the implemented SVM IP in order to classify a new test instance as melanoma or non-melanoma. Interestingly, any other trained SVM model with the same size of the implemented model can be easily implemented and run on Zynq by simply loading required data into the three main files. Therefore, the proposed system is promising for gaining scalability, flexibility, and adaptability.

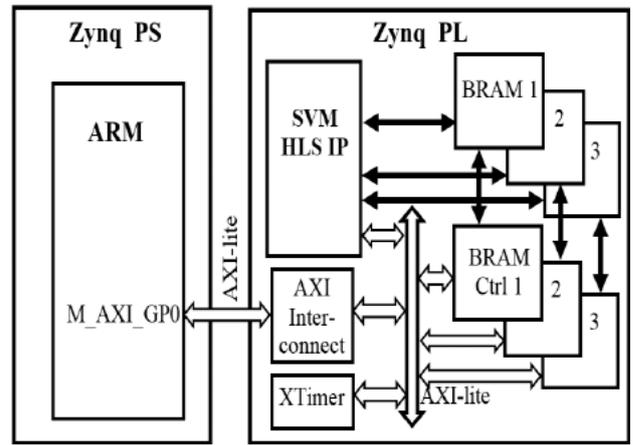

Figure 3. The proposed system on Zynq SoC.

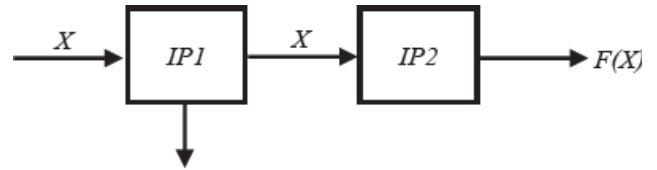

Figure 4. The proposed 2-stages cascaded classifier.

**C. Proposed Cascaded SVM Classification**

The proposed system is a scalable IP-based design that is easily extended to form a multi-core architecture by adding more SVM IPs in a single device/SoC that could be applied as a cascaded classification. The cascaded classification architecture contains multi classification stages (classifiers) that are designed in a cascading structure for accelerating the classification task. The early stages of the cascade are based on simple and low-complex classifiers, where the majority of data are classified and finalized (as melanoma/positive samples), while very few data are passed to the latter stages of more complex classifiers to be finally classified and verified.

The proposed SVM HLS IP is simplified to be used in the cascade. A new IP is proposed based on the designed IP in Fig. 2, where the first block is pre-calculated (offline on software) to store the value of the accumulated vector $Z$ (depends on trained SVM model data) in the IP as an internal memory. So, the new IP has a single input interface $X$ (assigned to the control bus instead of a BRAM interface) and consists of the two blocks, distance calculation and classification decision. The new IP is exploited to build a 2-stages cascaded classifier as in Fig. 4 as a case study, where the first SVM IP (denoted as *IP 1*) is a melanoma-sensitive classifier (61 SVs) and the second IP (denoted as *IP2*) is a non-melanoma-sensitive classifier with bigger size (139 SVs). This proposed architecture verifies the non-melanoma samples by passing input $X$ through the second classifier in order to reduce the number of false negative and risk of inaccurate diagnosis that harm patients. However, the melanoma results from the first stage should be verified by a skin cancer specialist for further diagnosis and medication.

Similarly, the two IPs are first implemented by the HLS tool to be integrated in a single system similar to the proposed in Fig. 4, where the two IPs are implemented in the Zynq PL and connected to the ARM processor. Also, an application is developed in the SDK tool to run the implemented cascaded classifier for validating online melanoma classification on Zynq SoC.

This proposed IP-based multi-core architecture as a 2-stages cascaded classification system is scalable, and could be easily extended in the future to implement a multi-stage system for higher classification accuracy and acceleration. Other scenarios would be applied easily to the cascaded architecture to comply with different applications needs and different sizes.

## 3. EXPERIMENTAL RESULTS AND DISCUSSION

We use Xilinx Vivado 2016.1 Design Suite to design and implement our proposed system on the Xilinx XC7Z020CLG484-1 target device of the Zynq-7 ZC702 Evaluation Board. The "SVM-Light" is implemented for our SVM IP for melanoma detection. SVM-light is available in C language, which is widely utilized for classification in numerous areas and applications [7]. The modern UltraFast HLS design methodology is used to implement an SVM IP on the Zynq SoC, based on the binary classification algorithm and C/C++ code of the SVM-Light. For all used data, float data type is defined in the C code and mapped in hardware to the standard single-precision floating-point format.

By using the available SVM-Light windows application, the training phase is utilized for generating our binary SVM model offline, using the default parameters and the linear kernel function. A dataset of 356 clinical images (168 melanoma and 188 benign images) collected from available web resources was used as input to the developed images preprocessing and feature extraction algorithms [2]. The extracted features from the image dataset were used in the training phase to generate a trained model for melanoma detection. Each image includes only a color image of one mole with the diameter of 6 mm or greater. Some selected pre-processing algorithms are applied first to the lesion images for hair artifact removal and then they are manually cropped and resized to form unified images of 512×512 pixels. Next, a lesion segmentation or border detection algorithm (interactive object recognition) is employed for background removal. Finally, feature extraction schemes based on HSV color channels are applied to the images for generating the features extracted dataset to be used for the SVM training (maximum number of features equals 27) [2]. The cross-validation technique is exploited in the training phase to achieve acceptable accuracy rates. Then, the generated trained model with all data specified from the training phase is extracted to be used for the hardware implementation of our embedded classification system.

Two SVM models/IPs are implemented based on the proposed design on Zynq SoC after extracting required data from the training phase. First, a small-scale model "*model 1*" of 61 SVs and 27 features is implemented, which is trained with a part of the available melanoma features dataset (100 melanoma and 44 benign instances). The trained model has a significant accuracy of 97.92% for melanoma detection, which shows low values of hardware resources utilization with only 2 watts of total power consumption for Zynq implementation (see Table I). The device statically dissipates 8% (0.17 W) of the total power consumption, whereas the remainder *92%* (1.89 W) is consumed by the dynamic activity, where mostly dissipated by the Zynq PS component (*74%* (1.4 W) of total dynamic power) compared to other on-chip components. Interestingly, a significant acceleration factor of *26* is achieved compared to running an equivalent classification software processing on the embedded ARM processor at the Zynq PS part. A processing time of 11.46 μs (2865 clock cycles) is achieved by using a high operating frequency of 250 MHz offered on the modern Zynq platform, while 309.36 μs (77340 clock cycles) is achieved for the ARM processing at the same frequency.

Another large-scale model "*model 2*" is implemented that is trained using the whole melanoma dataset (168 melanoma and 188 benign instances) and gives an SVM model of 248 SVs and 27 features with 80.85% accuracy. Similarly, low values of resources utilization and power consumption (2.65 W) (see Table I) are achieved with an acceleration factor of **32x** and 39.3 μs processing time, while 1.24 ms are required by the ARM processor. Consequently, the proposed design is capable of meeting the challenging constraints of embedded systems of high performance, low area, cost and power in addition to flexibility and scalability, which is so promising to realize a low-cost handheld device for efficient melanoma detection.

The proposed 2-stages cascaded classifier architecture "*cascaded model*" is implemented with two trained SVM models dividing the given dataset to generate a melanoma-sensitive model as *IP 1* with 61 SVs and 97.92% accuracy (100 melanoma and 44 benign dataset) and a benign-sensitive model as *IP 2* with 139 SVs and 72.51% accuracy (67 melanoma and 144 benign dataset). By implementing the cascaded classification structure with the simplified design of the SVM IPs, lower values of the area (except the total DSPs is doubled) and power consumption (1.74 W) are achieved over a single full SVM classifier, while improving the classification accuracy and speed (1.8 us) in addition to diagnosis verification. Also, an acceleration factor of *5* is achieved compared to software implementation of the cascaded classifier on ARM processor.

The proposed hardware design successfully extends our previous hardware/software co-design [6] to implement the whole SVM classification function onto the Zynq PL(FPGA), while some extra resources are utilized (same number of DSPs) with very little increase in power dissipation (< 1 W). To the best of our knowledge, this is the first cascaded SVM classification system for melanoma detection on Zynq SoC with optimized hardware results and high classification accuracy.

The implemented SVM models and the cascaded model are validated by running the developed applications (test benches) on the Zynq SoC for online melanoma classification. Some instances are tested to be correctly classified by all implemented models as melanoma/benign class with exactly the same classification result of the software application. Accordingly, the classification accuracy level is preserved without any loss from the hardware implementation, in contrast to other reported implementations in the literature [8–11]. Interestingly, our implemented SVM models show high acceptable detection accuracy that could be used and applied in real life. More instances would be tested in the future in order to validate the classification accuracy rate of the hardware implemented classifiers, as well as calculating sensitivity and specificity rates.

Table I. IMPLEMNTATION RESULTS ON ZYNQ SOC

| Models | Resources Utilization (%) | | | | | P (W) |
|---|---|---|---|---|---|---|
| | Slices (106400) | LUT (53200) | LUT-RAM (17400) | BRAM (140) | DSP (220) | |
| Model 1 | 10874 (10%) | 7218 (14%) | 874 (5%) | 48 (34%) | 5 (2%) | 2.06 |
| Model 2 | 30006 (28%) | 17506 (33%) | 2873 (17%) | 48 (34%) | 5 (2%) | 2.65 |
| Cascaded Model | 4304 (4%) | 3414 (6%) | 215 (1%) | 2 (1%) | 10 (5%) | 1.74 |

In addition, the implemented SVM classifiers are remarkably accelerated onto hardware to realize real-time embedded systems, which also outperform some existing implementations regarding processing time [8], [9], [12]–[13][14]. Moreover, the implemented models are realized onto the recent FPGA technology of the modern Zynq SoC, in contrast to most existing implementations that used old versions of FPGAs. The Zynq implemented systems achieve significantly low resources utilization and power consumption, which demonstrate lesser values than other implementations in the literature [9–15].

Finally, our implemented models on the recent hybrid Zynq SoC platform achieve optimized results for the hardware resources utilization, power consumption, detection speed and processing time with high classification accuracy rates using real data of melanoma detection.

## 4. CONCLUSION

Our Zynq-based embedded systems implementation using UltraFast HLS design methodology is the first FPGA-based SVM classifier that targets melanoma classification. In addition, our implemented systems successfully overcome most challenges exit in the literature of meeting critical embedded systems constraints of high performance, flexibility, scalability, and low levels of area, cost, and power consumption, while reaching reliable effective classification with high accuracy. The presented scalable IP-based multicore (cascaded) architecture could be easily extended in the future for a multi-stages system realization and also could be applied as a multi-class or ensemble classification and for various scenarios. Finally, the implemented classifier is feasible to be embedded in the future within a fast low-cost handheld medical scanning device dedicated for melanoma detection or any other applications.

## REFERENCES

[1] H. M. Hussain, K. Benkrid, and H. Seker, "The Role of FPGAs as High Performance Computing Solution to Bioinformatics and Computational Biology Data," AIHLS2013, p. 102, 2013.

[2] P. Sabouri, H. GholamHosseini, T. Larsson, and J. Collins, "A Cascade Classifier for Diagnosis of Melanoma in Clinical Images," in 36th Annual International Conference of the IEEE Engineering in Medicine and Biology Society (EMBC), 2014, pp. 6748-6751.

[3] S. M. Afifi, H. GholamHosseini, and R. Sinha, "Hardware Implementations of SVM on FPGA: A State-of-the-Art Review of Current Practice," International Journal of Innovative Science, Engineering &Technology (IJISET), vol. 2, pp. 733- 752, 2015.

[4] Zynq-7000 All Programmable SoC. Available: http://www.xilinx.com/products/silicon-devices/soc/zynq-7000.html

[5] Vivado High-Level Synthesis. Available: http://www.xilinx.com/products/design-tools/vivado/integration/esl-design.html

[6] S. Afifi, H. GholamHosseini, and R. Sinha, "Hardware Acceleration of SVM-Based Classifier for Melanoma Images," in Image and Video Technology – PSIVT 2015 Workshops: RV 2015, GPID 2013, VG 2015, EO4AS 2015, MCBMIIA 2015, and VSWS 2015, Auckland, New Zealand, November 23-27, 2015. Revised Selected Papers, F. Huang and A. Sugimoto, Eds., ed Cham: Springer International Publishing, 2016, pp. 235-245.

[7] T. Joachims, Making large-Scale SVM Learning Practical. Advances In Kernel Methods: Support Vector Learning, B. Schölkopf and C. Burges and A. Smola (ed.): MIT Press, 1999.

[8] M. Cutajar, E. Gatt, I. Grech, O. Casha, and J. Micallef, "Hardware-based Support Vector Machine for Phoneme Classification," in IEEE EuroCon 2013, 2013, pp. 1701-1708.

[9] M. Qasaimeh, A. Sagahyroon, and T. Shanableh, "FPGA-Based Parallel Hardware Architecture for Real-Time Image Classification," IEEE Transactions on Computational Imaging, vol. 1, pp. 56-70, 2015.

[10] C. Kyrkou, T. Theocharides, and C.-S. Bouganis, "An Embedded Hardware-Efficient Architecture for Real-Time Cascade Support Vector Machine Classification," in 2013 International Conference on Embedded Computer Systems: Architectures, Modeling, and Simulation (SAMOS XIII), 2013, pp. 129-136.

[11] C. Kyrkou, C. Bouganis, T. Theocharides, and M. M. Polycarpou, "Embedded Hardware-Efficient Real-Time Classification With Cascade Support Vector Machines," IEEE Transactions on Neural Networks and Learning Systems, 2015.

[12] M. Pietron, M. Wielgosz, D. Zurek, E. Jamro, and K. Wiatr, "Comparison of GPU And FPGA Implementation of SVM Algorithm for Fast Image Segmentation," in Architecture of Computing Systems–ARCS 2013, ed: Springer, 2013, pp. 292- 302.

[13] C. Kyrkou and T. Theocharides, "SCoPE: Towards a Systolic Array for SVM Object Detection," IEEE Embedded Systems Letters, vol. 1, pp. 46-49, 2009.

[14] M. Berberich and K. Doll, "Highly Flexible FPGA-Architecture of a Support Vector Machine," in MPC-Workshop 45, 2014, pp. 25-32.

[15] C. Kyrkou and T. Theocharides, "A Parallel Hardware Architecture for Real-Time Object Detection with Support Vector Machines," IEEE Transactions on Computers, vol. 61, pp. 831-842, 2012.